# Structural study on hole-doped superconductors Pr$_{1-x}$Sr$_x$FeAsO

J Ju[1], Z Li[1], G Mu[2], H-H Wen[2], K Sato[3], M Watahiki[3], G Li[4] and K Tanigaki[1,3]

[1]*World Premier International Research Center, Tohoku University, Sendai, 980-8578, Japan*

[2]*Institute of Physics, Chinese Academy of Sciences, Beijing, 100190, China*

[3]*Department of Physics, Graduate school of Science, Tohoku University, Sendai, 980-8578, Japan*

[4]*College of Chemistry and Molecular Engineering, Peking University, Beijing, 100871, China*

jujing@sspns.phys.tohoku.ac.jp

## Abstract

The structural details in Pr$_{1-x}$Sr$_x$FeAsO (1111) superconducting system are analyzed using data obtained from synchrotron X-ray diffraction and the structural parameters are carefully studied as the system is moving from non-superconducting to hole-doped superconducting with the Sr concentration. Superconductivity emerges when the Sr doping amount reaches 0.221. The linear increase of the lattice constants proves that Sr is successfully introduced into the system and its concentration can accurately be determined by the electron density analyses. The evolution of structural parameters with Sr concentration in Pr$_{1-x}$Sr$_x$FeAsO and their comparison to other similar structural parameters of the related Fe-based superconductors suggest that the interlayer space between the conducting As-Fe-As layer and the insulating Pr-O-Pr layer is important for improving $T_c$ in the hole-doped (1111) superconductors, which seems to be different from electron-doped systems.

*PACS 61.05.cp, 74.62.Bf, 74.62.Dh*

**Introduction**

Soon after the discovery of iron-based oxypnictide superconductors [1, 2], several structural families have been established to date, and now their structure is named in similarity to cuprates as (1111), (122), (111) and (11) [3]-[8]. The highest superconducting transition temperature ($T_c$) was achieved in an electron-doped system as 56.5 K in $Ca_{1-x}Nd_xFeAsF$ [9], and 55K in $SmFeAsO_{1-x}F_x$ [10], which is only surpassed by high $T_c$ cuprates.

A lot of attention has been paid so far on the electron-doped (1111) superconductors and their structural details have been studied by X-ray and neutron diffraction experiments. Physical properties have also been clarified from the viewpoint of electronic transport as well as magnetic measurements, and the phase diagrams seem to be established. Recently, the symmetry of $FeAs_4$ tetrahedron is proposed to be one of the most essential parameters in controlling $T_c$ in the case of electron-doped (1111) systems from the systematic studies on $REFeAsO_{1-x}$ (RE = La and Nd) and $CeFeAsO_{1-x}F_x$ superconductors [11]-[13], *i.e.*, the $T_c$ has a maximum when the angle is close to the 109°28'.

Although the electron-doped (1111) superconductors have triggered a lot of attention towards higher transition temperatures by applying various rare-earth metals (RE = La-Nd, Sm, Gd, Tb, Dy, Ho), a small number of hole-doped superconductors adopting (1111) structure are reported to be possible in $La_{1-x}Sr_xFeAsO$ [4, 14, 15], $Pr_{1-x}Sr_xFeAsO$ [16], $Nd_{1-x}Sr_xFeAsO$ [17], $La_{1-x}Sr_xNiAsO$ [18] and $Gd_{0.9}Sr_{0.1}ONiBi$ [19] and their structural details have not been established yet. The highest $T_c$ in these hole-doped systems is 25 K for $La_{1-x}Sr_xFeAsO$ and the situation seems to be different from what had been experienced in the electron-doped (1111) systems. Especially, there are a lot of structural ambiguities, such as the actual $Sr^{2+}$ doping concentration, the influence of $Sr^{2+}$ on their structural parameters although these are very important in order to find the crucial factors for tuning superconductivity. Therefore, it is of great significance to prepare a high quality hole-doped (1111) phase and to carry out careful structural studies on them, in order to understand the relationship between structure and $T_c$ and moreover to search for further possibility in elevating $T_c$. For this experimentally-based reason, we have focused on the high quality sample preparation of $Pr_{1-x}Sr_xFeAsO$ series. This paper will present their structural parameters by using high resolution synchrotron radiation and make comparison of the relationships between structure and $T_c$ both for electron-doped and hole-doped systems.

**Experimental**

$Pr_{1-x}Sr_xFeAsO$ samples with nominal Sr doping amount *x* from 0.10 to 0.30 were successfully synthesized by using a two-step solid state reaction method. Firstly, PrAs and SrAs precursors were obtained by reacting Pr chips (purity 99.95%) and Sr chips (purity 99.5%) with As powder (purity 99.99%) in 1:1 ratio. The mixtures were ground and pressed into pellets. Then they were sealed in evacuated quartz tubes followed by heating at 700°C for 10 hours. Secondly, the precursors were smashed and ground together with Fe powder (purity 99.99%) and $Fe_2O_3$ powder (purity 99.99%) in stoichiometry as the formula $Pr_{1-x}Sr_xFeAsO$. Those samples were pressed into pellets and sealed in evacuated quartz tubes and heated at about 940 °C for 5 hours, followed by an annealing at 1150 °C for 48 hours. Then it was cooled down slowly to room temperature.

Synchrotron powder XRD experiments were performed on a large Debye–Scherrer camera installed at SPring-8 beam line BL02B2 by using an imaging plate as the detector. The wavelength of the X-ray was determined as 0.60261 Å by using $CeO_2$ as the reference. Glass capillaries with an inner diameter of 0.3 mm were used to hold the powder samples in order to eliminate the preferred orientation. The Rietveld refinements were carried out using GSAS in angle range of 2-70° with an increment of 0.01° [20].

Resistivity was measured by employing a standard four-probe method using silver paste for contact. Measurements were carried out with a Quantum Design Physical Property Measurement System (PPMS). DC magnetic susceptibility measurements were performed in a Quantum Design superconducting quantum interference device (MPMS-XL).

**Results**

$Pr_{1-x}Sr_xFeAsO$ ($x$ = 0.1-0.3) polycrystalline samples with high purity were successfully synthesized using the two-step method aforementioned. We used high resolution synchrotron facility at Spring-8 to study the structural details of these samples. The Rietveld refinement results are listed in table 1 and a typical refinement pattern is shown in figure 1. All these samples adopt the tetragonal symmetry with the space group of $P4/nmm$ (129) at room temperature. Some samples have small amount of $Pr_2O_3$ and FeAs as impurities, which hinder the decrease of $R_p$ and $R_{wp}$. The structure of $Pr_{1-x}Sr_xFeAsO$ consists of interleaved two-dimensional FeAs and (Pr,Sr)O layers, as shown in the inset of figure 1. The FeAs layer is a conducting layer with Fe in four-fold coordination forming a $FeAs_4$-tetrahedron, whereas the (Pr,Sr)O layer is insulating and providing charge carriers to the conducting layer. Pr atoms are coordinated with four O atoms in the (Pr,Sr)O layer, and also weakly bonded with four As atoms in the neighboring FeAs layer. In the (Pr,Sr)O layer, Pr is positively trivalent ($Pr^{3+}$) and O is negatively divalent ($O^{2-}$); the counterpart FeAs layer is negatively monovalent with positively divalent Fe ($Fe^{2+}$) and negatively trivalent As ($As^{3-}$). The substitution of $Pr^{3+}$ by $Sr^{2+}$ causes the (Pr/Sr) site positive (3-δ) charges, and accordingly the Fe site positive (2+δ) charge. The charge carriers are transferred through the Pr/Sr plane and the FeAs layer as indicated in figure 1. This suggests that the effect of Sr doping is to facilitate hole transfer to induce hole-doped superconductivity.

We first checked how much amount of Sr was introduced into the structure. In all the samples with nominal compositions of $Pr_{1-x}Sr_xFeAsO$ ($x$ = 0.1-0.3), the Sr atom occupies the same atomic position (*2c*) as Pr with a certain amount of occupancy, which is precisely determined by the electron density analysis at this site. The results show that the actual doping concentration varies from 0.157 to 0.314. These values deviate from the nominal compositions, but the increase trend is basically the same. The actual doping concentration is used for the following structural analyses. The atomic occupancies of Fe, As and O were confirmed to be nearly 100%.

Given the situation that ionic radius of $Sr^{2+}$ (1.18 Å) is much larger than that of $Pr^{3+}$ (0.99 Å) [21], it is reasonable to see a remarkable increase of the lattice constants with the increase of Sr doping concentration. Actually, both *a* and *c* increase linearly with $x$ as shown in figure 2, where $x$ is the real doped Sr amount ranging from 0 to 0.314. The data at $x$ = 0 was taken from the previous report [22]. The increase in *a* is about 0.01 Å (increasing ratio = 0.25%) and *c* axis increases more obviously with 0.06 Å (increasing ratio = 0.7%) with this doping. Consequently, the unit cell volume evolves with Sr concentration in a linear relationship $V = 136.411 + 4.217x$, which proves the actual introduction of Sr into the structure.

Temperature dependencies of electric resistivity under zero field and magnetic susceptibility measured with the zero-field-cooling mode under 20 Oe are shown in figure 3. The samples with doping amount lower than 0.221 do not show superconductivity, even though their lattice constants obviously increase. However the electric resistivity and the magnetic susceptibility measurements give a clear superconductivity transition for the samples with the doping amount of 0.221-0.314. The onset temperature of superconductivity ($T_{onset}$) is determined by the first derivative of the electric resistivity as depicted in the inset of figure 3(a). $T_{onset}$ varies from 13.5 K to 15.1 K, with a maximum of 15.1 K at $x$ = 0.221. In addition, the sample with $x$ = 0.221 also gives the largest superconducting volume fraction as shown in figure 3(b), indicating that the optimal doping concentration is around 0.221.

In order to study the structure change influenced by doping, we carried out Rietveld refinements in detail based on the synchrotron X-ray powder diffraction patterns. The refinement results are listed in table 1. Figure 4 summarizes the influence of Sr doping on the crystal structure of $Pr_{1-x}Sr_xFeAsO$. In the (Pr,Sr)O layer, one could see that the Pr-O bond distance slightly increases to an extent of 0.0025 Å in total, while the two O-Pr-O bond angles increase more obviously with $x$. The largest increase of the two bond angles happens by 0.4° and 0.2° at the $x$ = 0.221 stoichiometry when the reference is taken at $x$ = 0.157. We define the Pr-O-Pr and the As-Fe-As block distances as the vertical distance between the Pr atoms residing at the top and the bottom in the (Pr,Sr)O layer and between the As atoms in the FeAs layer, respectively (as shown in figure 1). The increase in the Pr-O bond distance is not sufficient to compensate the large increase in the two O-Pr-O band angles, which unambiguously can explain the reason of the Pr-O-Pr block shrinkage. Especially, a remarkable reduction in the block distance is observed to be 0.013 Å at $x = 0.221$, taking $x = 0.157$ as reference.

In the As-Fe-As block, one could see small changes in the Fe-As bond distance and the two Fe-As-Fe bond angles. The Fe-As bond distance increases monotonically up to 0.005 Å with $x$. Meanwhile the decrease of the two Fe-As-Fe bond angles is about 0.08° and 0.04°. Accordingly, the changes in the bond distance and angles synergistically expand the As-Fe-As block distance from 2.667 Å to 2.673 Å. To be noted, this expansion is less than a half of the change observed in the Pr-O-Pr block, which results in a remarkable increase of the interlayer distance.

**Discussions**

Some relatively larger $Sr^{2+}$ (1.18 Å) substitutes the smaller $Pr^{3+}$ (0.99 Å) at the same atomic site, and therefore it causes the expansion of the unit cell which was confirmed by X-ray analyses as shown in table 1 and figure 2. The unit cell shows a linear expansion and obeys the Vergard's law, giving the further experimental evidence that the replacement of Pr by Sr takes place. Moreover, the introduction of Sr modifies the crystal structure in a delicate way as summarized in figure 4. The Pr-O and Fe-As bond distances increase slightly with $x$ nevertheless the values are still comparable with other related compounds. The two Fe-As-Fe bond angles decrease with $x$, slowly moving toward the ideal values of the perfect $FeAs_4$ tetrahedron. The evolution of $T_c$ as a function of *M-Pn-M* bond angle is illustrated in figure 7(c) together with those reported in the electron-doped systems [13]. As being admitted in the electron-doped pnictide systems, the regular angle of the $MPn_4$ ($M$ = Fe or Ni, $Pn$ = As or P) tetrahedron is crucial for enhancement of $T_c$. The rich crystallographic information so far available in the electron-doped systems gives a clear tendency that the highest $T_c$ is achieved at the nearly perfect $MPn_4$ tetrahedral point. In contrast only three sets of detailed structural parameters are available in the hole-doped systems, *i.e.* $Nd_{1-x}Sr_xFeAsO$ [17], $Ba_{1-x}K_xFeAsO$ [6] and $Pr_{1-x}Sr_xFeAsO$ in the present work. Although one can see a similar tendency also for the hole-doped systems, *i.e.* the regular angle of the $FeAs_4$ tetrahedron helps to enhance $T_c$, the $T_c$ values are always much smaller than those in the electron-doped systems. This phenomenon implies that the Fe-As-Fe bond angle may not be regarded as only the unique parameter to tune the superconductivity in the hole-doped systems.

Different from the slight changes in the Pr-O and Fe-As bond distances as well as the Fe-As-Fe bond angles, the Sr doping greatly changes the two O-Pr-O bond angles and results in the shrinkage of the Pr-O-Pr block. As consequence, the Pr-As bond distance is largely increased. The Pr-As bond distance can give a measure of the interlayer space between the Pr-O-Pr insulating block and the As-Fe-As conducting block. Figure 5 summarizes the information on the shrinkage of the Pr-O-Pr block, the expansion of the As-Fe-As block and the increase of the Pr-As distance. The Pr-As distance increases greatly with $x$ and reaches a maximum of 0.034 Å compared with the parent PrFeAsO [22].

Figure 6 displays the influence of the Pr-As distance on $T_{onset}$, which was deduced from d$\rho$/d$T$ as given in figure 3(a). In the superconductivity region, $T_{onset}$ increases linearly with the Pr-As distance, indicating that the larger interlayer space favors to improve $T_c$. Actually, this feature is observed in the other two hole-doped systems as well. Figure 7(a) shows $T_c$ as a function of $M$-As bond distance in the three hole-doped pnictide superconductors, where $M$ is Ba/K for Ba$_{1-x}$K$_x$Fe$_2$As$_2$, Nd/Sr for Nd$_{1-x}$Sr$_x$FeAsO and Pr/Sr for the present system. The $M$-As bond distances in these compounds are calculated from the reported structural information. Clearly, superconductivity emerges when the interlayer space is expanded to a large extent in all these hole-doped superconductors. The situation can be compared to the different fact that the interlayer space is decreased in several electron-doped (1111) systems. For instance, the Ce-As distance in CeFeAsO$_{1-x}$F$_x$ decreases linearly from 3.33 to 3.28 Å with $x$ = 0 to 0.16 [12], which is believed to help to bring the Ce(O,F) charge transfer layer closer to the superconducting FeAs one, thereby facilitating electron carrier transfer. Figure 7(b) shows the evolution of $T_c$ as a function of the RE-As bond distance in REFeAsO$_{1-x}$F$_x$ (RE = Ce) and REFeAsO$_{1-y}$ (RE = La and Nd) systems [11]-[13]. The observed feature is opposite to the case discussed earlier in the hole-doped systems.

In order to understand the expansion of the interlayer space in the hole-doped systems, the valence state at the Pr site can be considered. We employed the bond-valence-sum (BVS) theory to evaluate the valence state of Pr [23], where each bond with a distance $r$ contributes to the valence $v = \exp[(d-r)/0.37]$ with $d$ as an empirical parameter. Considering the coordination of the Pr site, Pr is bonded with four O atoms in the Pr-O-Pr block and with the other four As atoms in the As-Fe-As block. In the non-superconducting compound with $x$ = 0.157, the four Pr-O bonds are estimated to contribute 2.392 to the Pr-BVS after correction for the Sr doping. The four Pr-As bonds, which are significantly longer, contribute 0.827 to the Pr-BVS. The total Pr-BVS 3.219 is in good agreement with the expected Pr valence. In the superconducting compound with $x$ = 0.221, the increase in the Pr-O and Pr-As bond distances reduces the Pr-BVS to 2.382 and 0.793 respectively, which gives a total Pr-BVS value of 3.175. These values clearly show that the increase of Pr-As bond distance plays an important role in the reduction of the Pr-BVS, and give further confirmation that the replacement of Sr decreases the charge at the Pr site. Considering the layer structure of Pr$_{1-x}$Sr$_x$FeAsO, Pr/Sr resides at the top and the bottom of the Pr-O-Pr block, while As resides at the top and the bottom of the As-Fe-As block. Therefore Coulombic interactions between the neighboring Pr-O-Pr and As-Fe-As block are weakened significantly by the Sr-doping so that the interlayer space can be expanded.

The expansion of the interlayer space in the hole-doped pnictide systems is reminiscent of the layer-structured metal nitrides, $\beta$-$M$NCl ($M$ = Zr and Hf) [24]. Upon expansion of the interlayer space by the intercalation of Li and tetrahydrofuran in this system, the higher transition temperatures are observed. Due to the limitation of the structural information for the hole-doped systems at the moment, we can not strictly describe how the interlayer expansion favors to increase $T_c$ in the hole-doped (1111) systems, but the present result will provide important suggestion that the larger interlayer space between the conducting layer and the carrier-providing insulating layer is one of the key parameters for improving $T_c$ in the hole-doped systems.

**Conclusion**

We successfully synthesized the hole-doped (1111) superconductors $Pr_{1-x}Sr_xFeAsO$. Superconductivity emerges when $x$ is larger than 0.221 with a large superconducting fraction. Careful comparison between the structural parameters and $T_c$ among other various Fe-based superconductors reveals that the interlayer space expands systematically in $Pr_{1-x}Sr_xFeAsO$ with increasing $T_c$. $T_c$ reaches its maximum when the interlayer space is the largest. This suggests that the interlayer space is one of the crucial parameters for achieving the higher $T_c$ in the hole-doped (1111) systems. This is a different trend from that encountered for the electron-doped (1111) systems.

**Acknowledgement**

This work was performed by a Grant-in-Aid from the Ministry of Education, Culture, Sports, Science, and Technology of Japan, No.18204030, 19014001, 18651075 and 18204032. This work was carried out under Grant-in-Aid for Scientific Research on Priority Areas "New Materials Science Using Regulated Nano Spaces-Strategy in Ubiquitous Elements" from the Ministry of Education, Culture, Sports, Science and Technology of Japan. The synchrotron radiation experiments were performed by the approval of the Japan Synchrotron Radiation Research Institute (JASRI) as Nanotechnology Support Project. This work was partially supported by Grants-in-Aid for Scientific Research from the Japan Society for the Promotion of Science (JSPS) (Grants No.P07025). The research is also partially supported by Tohoku university GCOE program.

**Reference**

1. Kamihara Y, Watanabe T, Hirano M and Hosono H 2008 *J. Am. Chem. Soc.* **130** 3296
2. Kamihara Y, Hiramatsu H, Hirano M, Kawamura R, Yanagi H, Kamiya T and Hosono H 2006 *J. Am. Chem. Soc.* **128** 10012
3. Chen G F, Li Z, Li G, Zhou J, Wu D, Dong J, Hu W Z, Zheng P, Chen Z J, Yuan H Q, Singleton J, Luo J L and Wang N L 2008 *Phys. Rev. Lett.* **101** 057007
4. Wen H H, Mu G, Fang L, Yang H and Zhu X 2008 *EPL* **82** 17009
5. Chen X H, Wu T, Wu G, Liu R H, Chen H and Fang D F 2008 *Nature* **453** 761
6. Rotter M, Tegel M and Johrendt D 2008 *PRL* **101** 107006
7. Hsu F C, Luo J Y, Yeh K W, Chen T K, Huang T W, Wu P M, Lee Y C, Huang Y L, Chu Y Y, Yan D C and Wu M K 2008 *Proceedings of the National Academy of Sciences* **105** 14262
8. Chu C W, Chen F, Gooch M, Guloy A M, Lorenz B, Lv B, Sasmal K, Tang Z J, Tapp J H and Xue Y Y arXiv:0902.0806
9. Cheng P, Shen B, Mu G, Zhu X, Han F, Zeng B and Wen H H 2009 Europhys. Lett. **85** 67003
10. Ren Z A, Lu W, Yang J, Yi W, Shen X L, Zheng C, Che G C, Dong X L, Sun L L, Zhou F and Zhao Z X 2008 *Chin. Phys. Lett.* **25** 2215
11. Cruz1 C, Huang Q, Lynn J W, Li J, Ratcliff W, Zarestky J L, Mook H A, Chen G F, Luo J L, Wang N L and Dai P 2008 *Nature* **453** 899
12. Zhao J, Huang Q, Cruz1 C, Li S, Lynn J W, Chen Y, Green M A, Chen G F, Li G, Li Z, Luo J L, Wang N L and Dai P 2008 *Nature materials* **7** 953
13. Lee C H, Iyo A, Eisaki H, Kito H, Fernandez-Diaz M T, Ito T, Kihou K, Matsuhata H, Braden M and Yamada K 2008 *Journal of the Physical Society of Japan* **77** 083704
14. Mu G, Fang L, Yang H, Zhu X, Cheng P and Wen H H 2008 *Proc. Int. Symp. Fe-Pnictide Superconductors, J. Phys. Soc. Jpn. Suppl. C* **77** 15
15. Wu G, Chen H, Xie Y L, Yan Y J, Wu T, Liu R H, Wang X F, Fang D F, Ying J J and Chen X H 2008 *Phys. Rev. B* **78** 092503
16. Mu G, Zeng B, Zhu X, Han F, Cheng P, Shen B and Wen H H 2009 *Phys. Rev. B* **79** 104501
17. Kasperkiewicz K, Bos J W G, Fitch A N, Prassides K and Margadonna S 2009 *Chem. Commun.* 707
18. Fang L, Yang H, Cheng P, Zhu X, Mu G and Wen H H 2008 *Phys. Rev. B* **78** 104528
19. Ge J, Cao S and Zhang J arXiv:0807.5045
20. Altomare A, Burla M C, Cascarano G, Giacovazzo G, Guagliardi A, Moliterni A G G and Polidori G 1995 *J. Appl. Crystallogr.* **28** 842
21. Shannon R D 1976 *Acta Cryst.* **A32** 751
22. Quebe P, Terbuchte L J and Jeitschko W 2000 *Journal of Alloys and Compounds* **302** 70

23. Brese N E and O'Keefee M 1991 *Acta Cryst.* **B47** 192
24. Yamanaka S, Hotehama K and Kawaji H 1998 *Nature* **392** 580

Table 1. Rietveld refinements results of $Pr_{1-x}Sr_xFeAsO$ by using synchrotron powder diffraction data at room temperature. Space group: *P4/nmm*. Atomic positions: Pr/Sr: *2c* (1/4, 1/4, *z*); Fe: *2b* (3/4, 1/4, 1/2); As: *2c* (1/4, 1/4, *z*) and O: *2a* (3/4, 1/4, 0). $U_{iso}$ is the isotropic displacement parameters constrained for the same chemical species. The occupation factors of $Sr^{2+}$ were refined and then fixed to appropriate values in the final refinement.

| Atom | | $x = 0.157(1)$ | $x = 0.162(1)$ | $x = 0.221(1)$ | $x = 0.262(1)$ | $x = 0.263(7)$ | $x = 0.278(3)$ | $x = 0.314(1)$ |
|---|---|---|---|---|---|---|---|---|
| | $a$ (Å) | 3.9845(4) | 3.9873(1) | 3.9936(1) | 3.9905(2) | 3.9902(9) | 3.9896(4) | 3.9902(9) |
| | $c$ (Å) | 8.6224(9) | 8.6190(2) | 8.6539(1) | 8.6486(7) | 8.6369(8) | 8.6473(8) | 8.6439(9) |
| | $V$ (Å) | 136.895(4) | 137.031(5) | 138.023(5) | 137.724(1) | 137.521(7) | 137.642(7) | 137.633(8) |
| Pr/Sr | $z$ | 0.1384(7) | 0.1380(5) | 0.1372(3) | 0.1380(3) | 0.1381(1) | 0.1381(4) | 0.1378(9) |
| | $U_{iso}$ | 0.0066(5) | 0.0064(3) | 0.0064(2) | 0.0064(8) | 0.0064(6) | 0.0067(1) | 0.0064(8) |
| Fe | $U_{iso}$ | 0.0079(4) | 0.0087(8) | 0.0102(1) | 0.0091(5) | 0.0089(9) | 0.0080(2) | 0.0086(4) |
| As | $z$ | 0.6546(7) | 0.6546(9) | 0.6543(5) | 0.6543(6) | 0.6545(2) | 0.6545(4) | 0.6546(1) |
| | $U_{iso}$ | 0.0060(1) | 0.0057(6) | 0.0055(4) | 0.0063(1) | 0.0062(1) | 0.0061(5) | 0.0065(3) |
| O | $U_{iso}$ | 0.0205(3) | 0.0146(8) | 0.0096(3) | 0.0183(6) | 0.0173(1) | 0.0206(2) | 0.0230(7) |
| | $Rp$ (%) | 3.13 | 3.84 | 3.46 | 3.12 | 3.81 | 3.09 | 4.54 |
| | $Rwp$ | 4.45 | 5.39 | 4.88 | 4.59 | 5.59 | 4.76 | 6.31 |

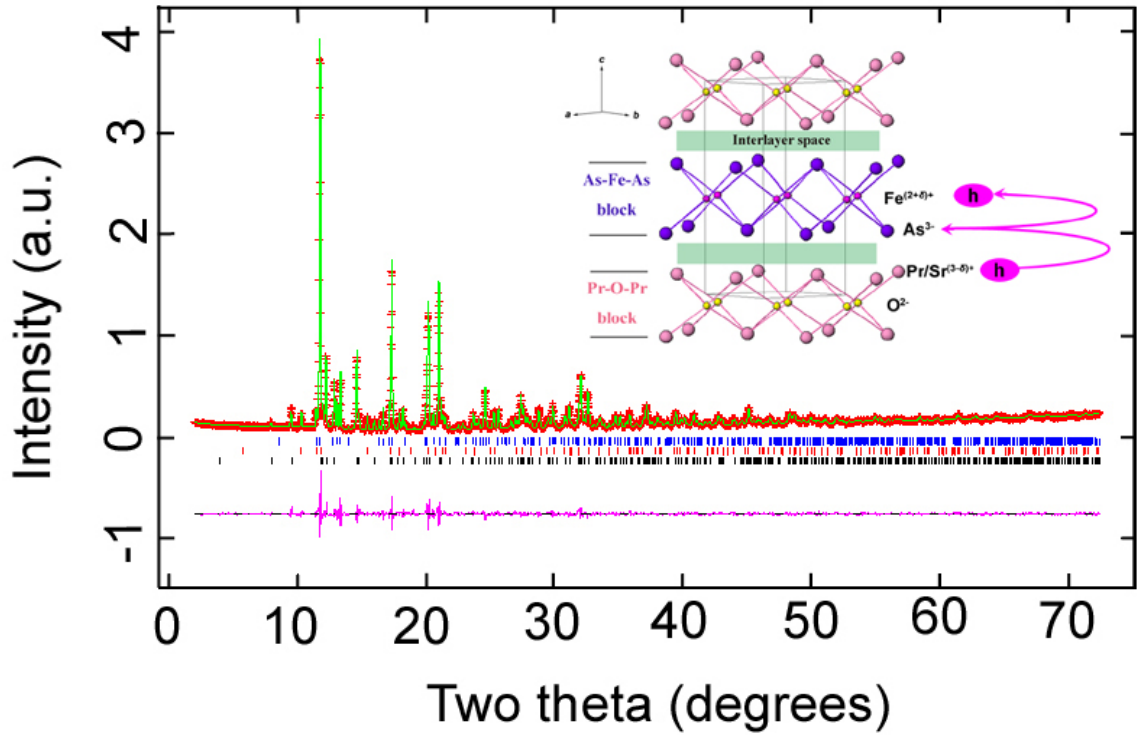

Figure 1. A typical observed (red crosses) and calculated (green solid line) X-ray powder diffraction pattern of $Pr_{1-x}Sr_xFeAsO$. Three rows of vertical bars show the calculated positions of Bragg reflections for FeAs (blue), $Pr_2O_3$ (red) and $Pr_{1-x}Sr_xFeAsO$ (black), respectively. The former two are impurities with a small amount. The purple solid line shown at the bottom of the figure is the differences between observations and calculations. The inset above the X-ray pattern is the schematic diagram defining the As-Fe-As block and the Pr-O-Pr block, and illustrating the process of hole transfer.

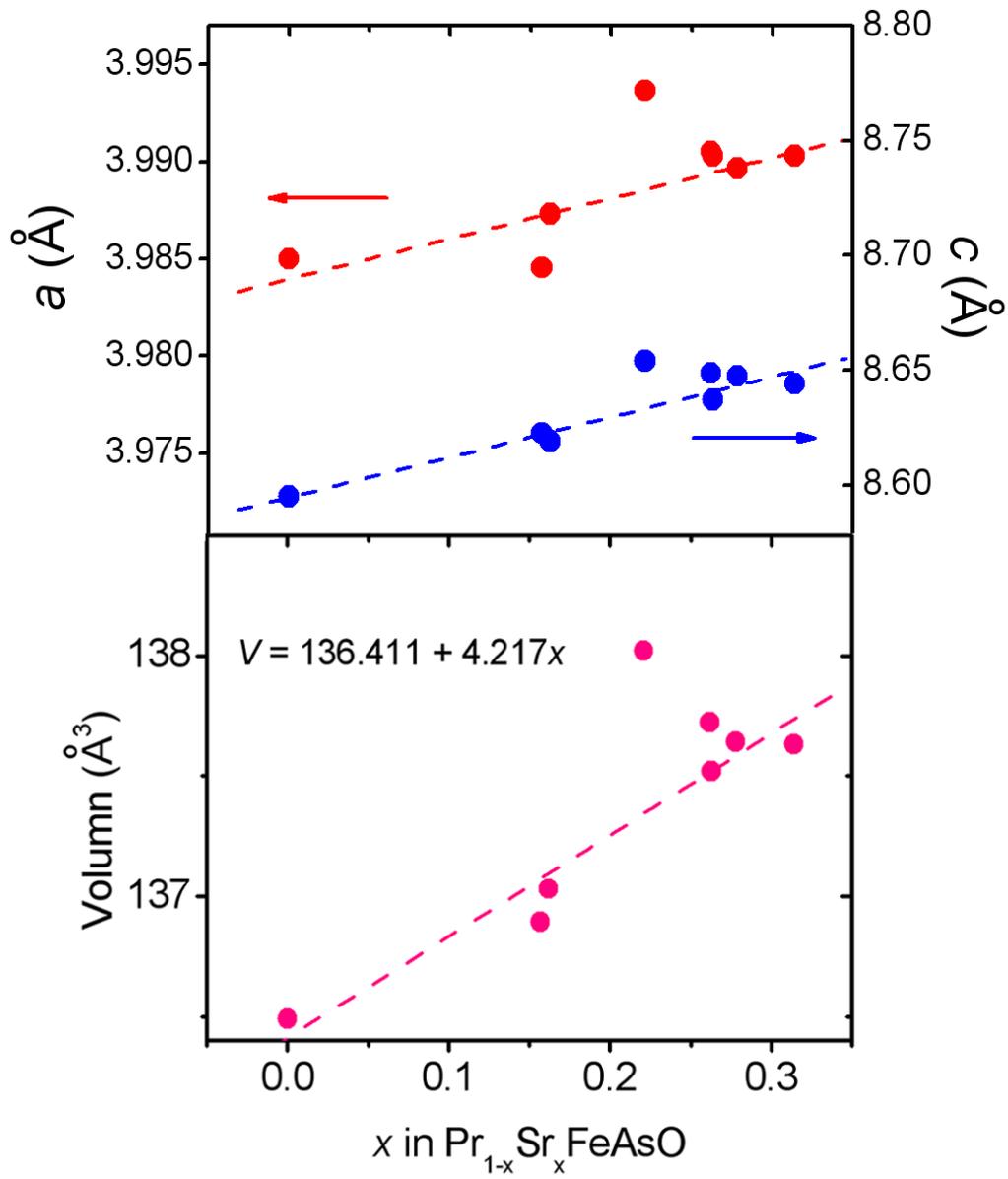

Figure 2. Lattice constants *vs* Sr doping amount in $Pr_{1-x}Sr_xFeAsO$. The dash lines are the linear simulation for various parameters. The data of non-doped PrFeAsO was cited from other report [22].

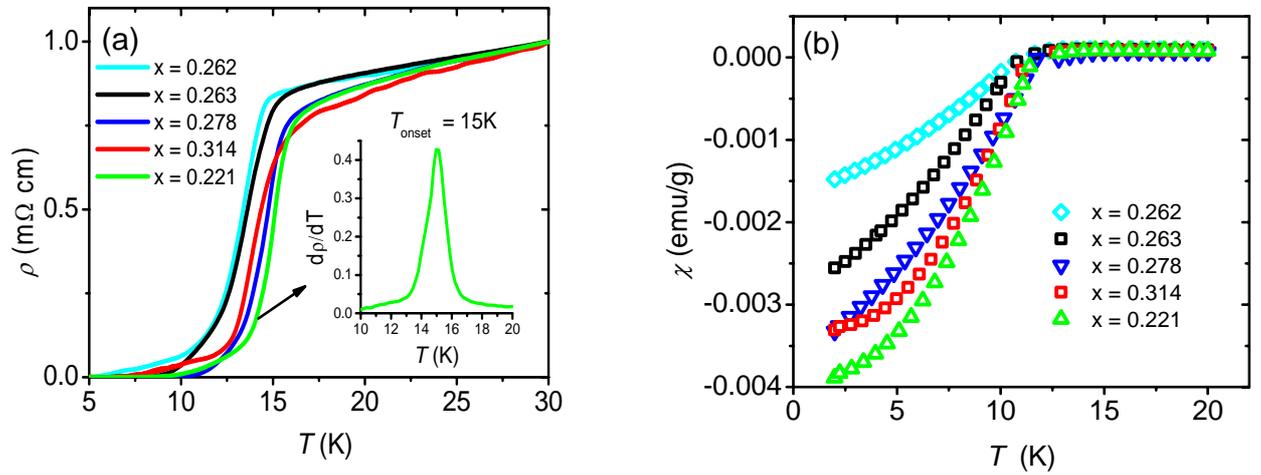

Figure 3. Temperature dependencies of electric resistivity (a) under zero field and magnetic susceptibility (b) measured with the zero-field-cooling mode under 20 Oe in $Pr_{1-x}Sr_xFeAsO$. The determination of the onset temperature of superconductivity ($T_{onset}$) is depicted in the inset of (a).

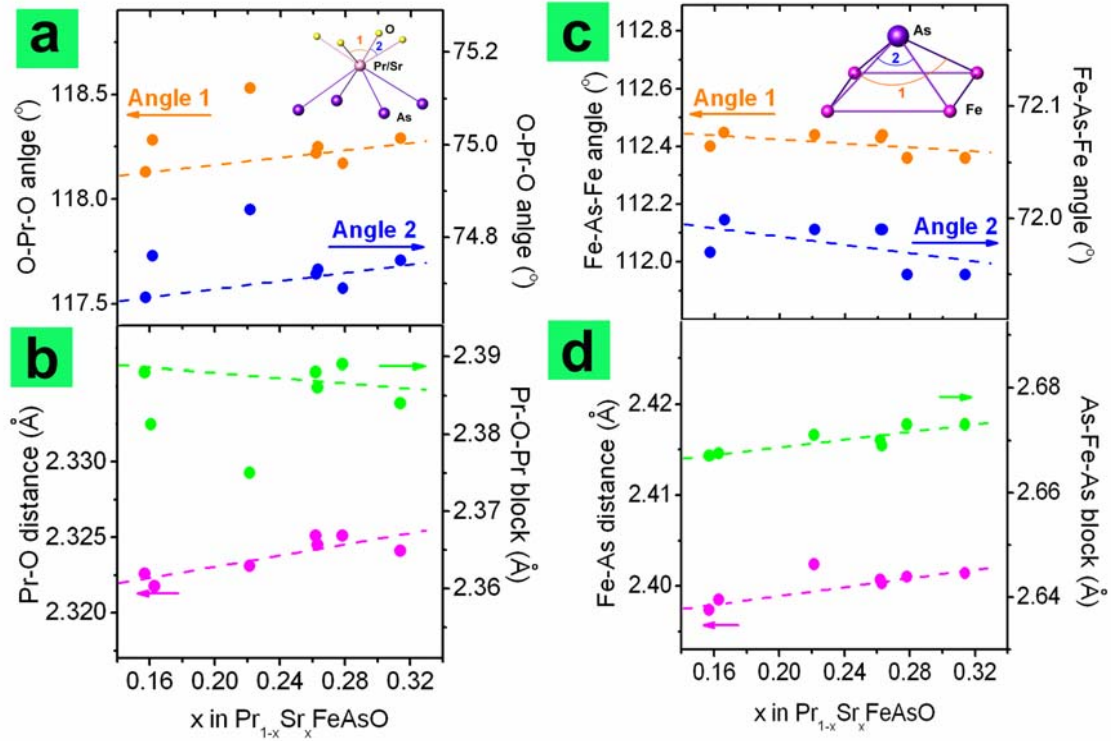

Figure 4. Structural evolution of $Pr_{1-x}Sr_xFeAsO$ as a function of Sr doping amount from the analysis of the synchrotron X-ray diffraction data. The atomic positions of $Pr_{1-x}Sr_xFeAsO$ are shown in Table 1 and the effect of Sr doping is to shrink the Pr-O-Pr block and to expand the As-Fe-As block, but with a total expansion of the interlayer space. *a*, O-Pr-O bond angles as a function of Sr doping. *b*, Pr-O bond distance and Pr-O-Pr block distance as a function of Sr doping. *c*, Fe-As-Fe bond angles as a function of Sr doping. *d*, Fe-As bond distance and As-Fe-As block distance as a function of Sr doping. The dash lines are the linear simulations for various parameters.

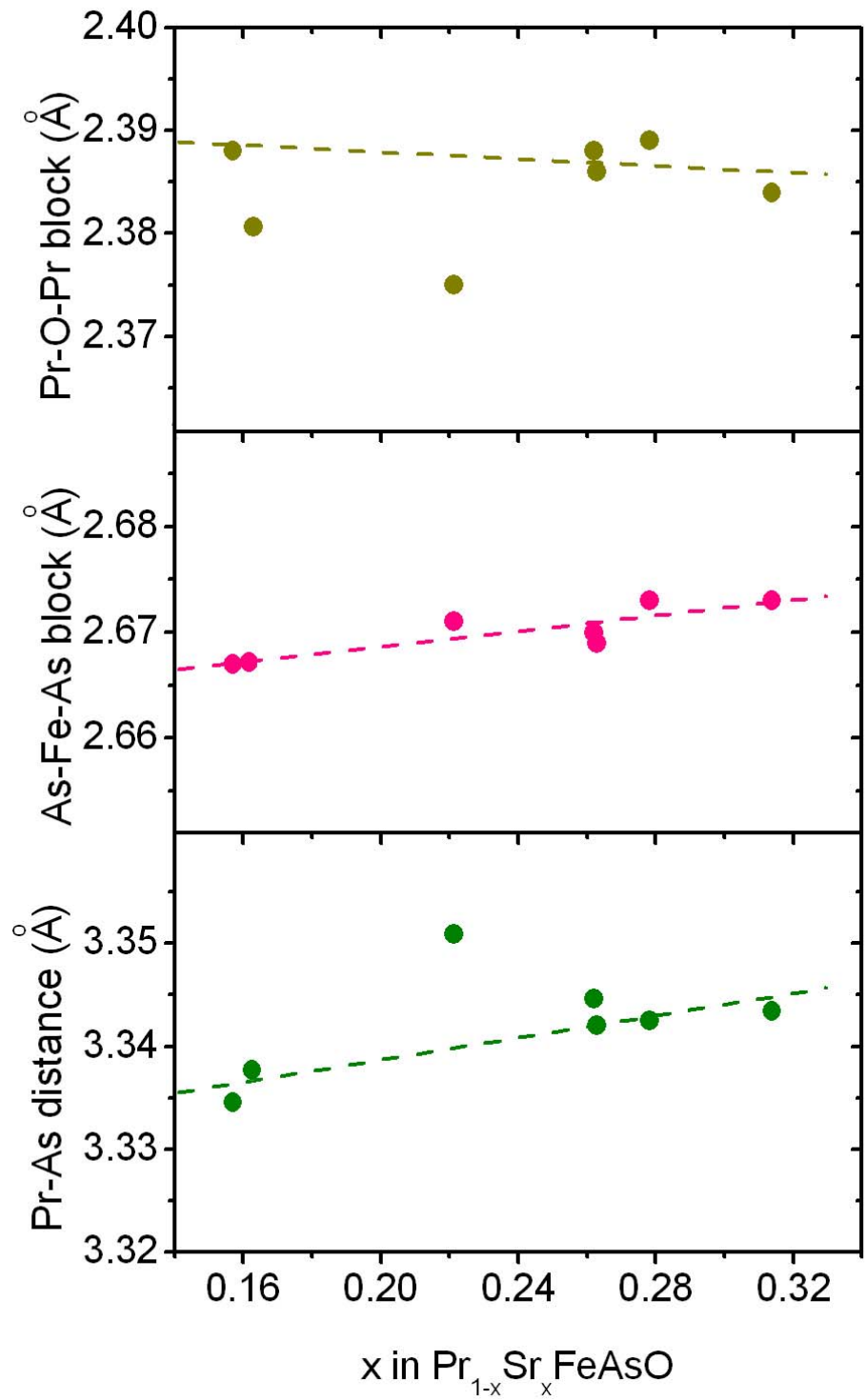

Figure 5. Evolution of Pr-O-Pr block, As-Fe-As block and Pr-As distance as a function of Sr doping in $Pr_{1-x}Sr_xFeAsO$. The dash lines are the linear simulations for various parameters.

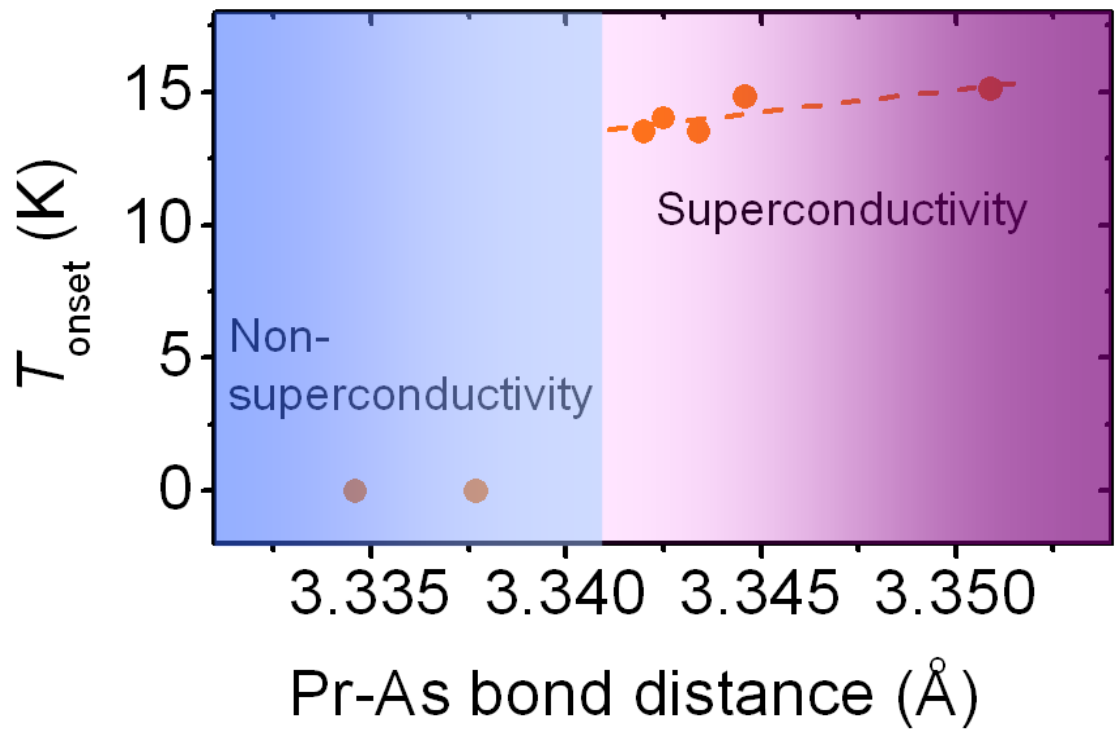

Figure 6. Impact of Pr-As bond distance on $T_{onset}$. $T_{onset}$ is deduced from $d\rho/dT$ as given in figure 3(a). In the superconductivity region, $T_{onset}$ increases linearly with the Pr-As distance, indicating the larger interlayer space favors to improve superconducting transition temperature.

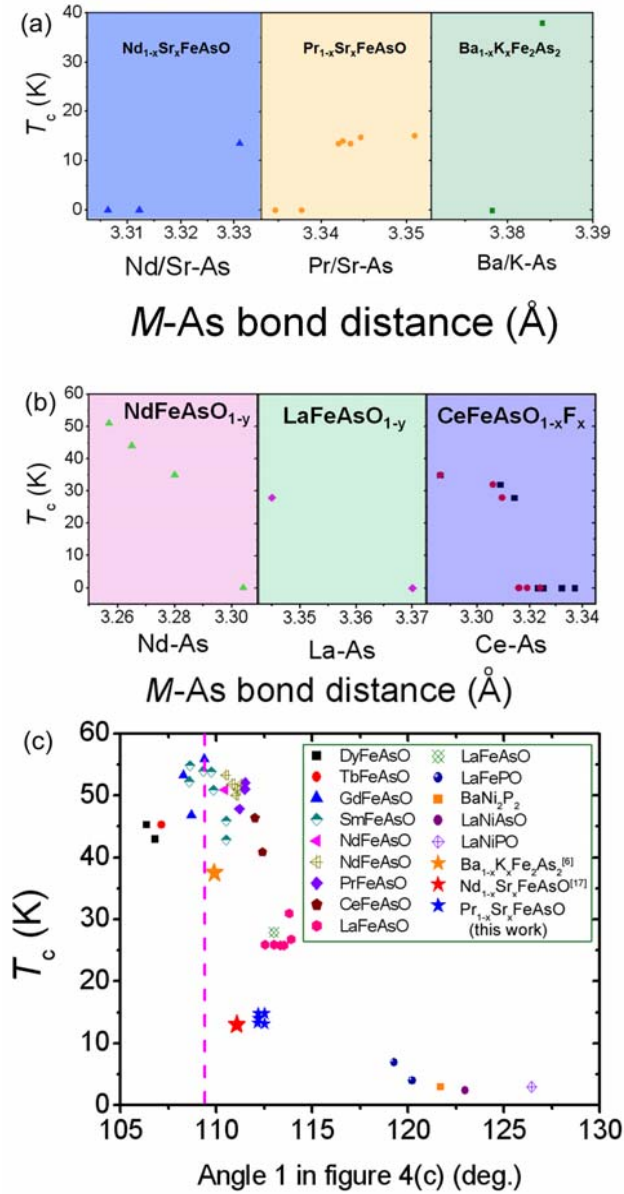

Figure 7. (a) $T_c$ as a function of *M*-As bond distance in various hole-doped superconductors, where *M* is Ba/K for $Ba_{1-x}K_xFe_2As_2$, Nd/Sr for $Nd_{1-x}Sr_xFeAsO$ and Pr/Sr for the present system, respectively. (b) $T_c$ as a function of *M*-As bond distance in various electron-doped systems, where *M* is Nd for $NdFeAsO_{1-y}$, La for $LaFeAsO_{1-y}$ and Ce for $CeFeAsO_{1-x}F_x$, respectively. Clearly, superconductivity emerges when the interlayer space is expanded in a large extent in the hole-doped systems. On the contrary, the interlayer space is shrunk in the electron-doped systems. (c) $T_c$ as a function of the bond angle 1 in various pnictide superconductors. The bond angle 1 was defined as the larger angle between transition metal and pnictide as shown in figure 4(c). Part of these data was taken from ref. [13]. The angle dependence of $T_c$ is largely weakened in the hole-doped systems, as compared with the electron-doped systems.